\documentclass[10pt]{wlscirep}
\usepackage[utf8]{inputenc}
\usepackage{float}
\usepackage[T1]{fontenc}
\usepackage[draft]{changes}
\usepackage{enumitem}
\usepackage{comment}
\usepackage{chngcntr}
\newcommand{\citep}{\cite} %

\usepackage{xcolor}
\definecolor{stanfordred}{RGB}{140,21,21}

\title{
What's documented in AI? Systematic Analysis of 32K AI Model Cards
}

\author[1*]{Weixin Liang}
\author[2*]{Nazneen Rajani} %
\author[3*]{Xinyu Yang}
\author[2]{Ezinwanne Ozoani} %
\author[4]{Eric Wu} %
\author[5]{Yiqun Chen} %
\author[6]{Daniel Scott Smith} %
\author[1,4,5+]{James Zou}

\affil[1]{Department of Computer Science, Stanford University, Stanford, CA, USA}
\affil[2]{Hugging Face, Inc.}
\affil[3]{Department of Information Science, Cornell University, Ithaca, NY, USA}
\affil[4]{Department of Electrical Engineering, Stanford University, Stanford, CA, USA}
\affil[5]{Department of Biomedical Data Science, Stanford University, Stanford, CA, USA}
\affil[6]{Graduate School of Education, Stanford University, Stanford, CA, USA}
\affil[+]{Correspondence should be addressed to: jamesz@stanford.edu}
\affil[*]{these authors contributed equally to this work}

\begin{abstract}

The rapid proliferation of AI models has underscored the importance of thorough documentation, as it enables users to understand, trust, and effectively utilize these models in various applications. Although developers are encouraged to produce model cards, it's not clear how much information or what information these cards contain. In this study, we conduct a comprehensive analysis of 32,111 AI model documentations on Hugging Face, a leading platform for distributing and deploying AI models. Our investigation sheds light on the prevailing model card documentation practices. Most of the AI models with substantial downloads provide model cards, though the cards have uneven informativeness. We find that sections addressing environmental impact, limitations, and evaluation exhibit the lowest filled-out rates, while the training section is the most consistently filled-out. We analyze the content of each section to characterize practitioners' priorities. Interestingly, there are substantial discussions of data, sometimes with equal or even greater emphasis than the model itself. To evaluate the impact of model cards, we conducted an intervention study by adding detailed model cards to 42 popular models which had no or sparse model cards previously. We find that adding model cards is moderately correlated with an increase weekly download rates. Our study opens up a new perspective for analyzing community norms and practices for model documentation through large-scale data science and linguistics analysis.

\end{abstract}
\begin{document}
\maketitle
\thispagestyle{empty}

\section*{Introduction}

The rising prevalence of AI models in various sectors has underscored the necessity of comprehensive model documentation~\cite{SWANSON20231772,Liang2022AdvancesCA,arrieta2020explainable}. As these models grow in complexity, their inner workings can seem increasingly obscure to those without specialized knowledge~\cite{molnar2020interpretable,shen2021value,seifert2019towards}. This creates a critical need for accurate and comprehensive documentation~\cite{modelcard,bender2021dangers}. Effective model documentation serves as a vital communication bridge between developers and users, offering explicit guidance on the model's functionality, from its inputs and outputs to its range of applications~\cite{FactSheets,seifert2019towards}. Importantly, it reveals potential biases, errors, and limitations inherent in the model~\cite{mehrabi2021survey}. This focus on transparency cultivates trust among users, a crucial component in fields where model output has far-reaching consequences, such as in healthcare, finance, and law enforcement~\cite{he2023blinded}. Ultimately, model documentation serves as a pivotal tool for improving model utility and credibility across a range of stakeholders~\cite{transparency1,raji2019ml,Accountability1, Accountability2,kulesza2015principles}.

Model cards have emerged as the standard approach to document AI models~\cite{modelcard,huggingface2021modelcard}, inspired by the concept of food nutrition labels~\cite{holland2018dataset} and datasheets~\cite{datasheets} in the electronics industry. They are documents that provide essential information about a model in a standardized, easy-to-understand format. At the core of these cards are sections detailing model training and validation procedures, intended uses and potential limitations such as bias and fairness analysis, and usage guidance. Compared to other documentation formats such as academic papers or technical reports, model cards are increasingly becoming a preferred reference for practitioners in the AI community for a number of reasons. First, they offer more concise, relevant, and easily understandable information about AI models, rendering them more accessible. Another important aspect of model cards is their up-to-date nature, as they can be frequently updated to reflect any changes, improvements, or new findings about the AI model. In contrast, academic papers, once published, may not be updated as regularly, which could result in outdated information. Additionally, many popular model repositories, especially those originating from industry or open-source enthusiasts, don't have accompanying academic papers or technical reports~\cite{databricks-dolly,vicuna2023}. This further accentuates the indispensable role of model cards as a comprehensive, streamlined, and informative communication mechanism within the AI ecosystem.

While many model cards are being created by researchers and developers~\cite{modelcard,FactSheets,shen2021value,seifert2019towards,bender2018data,bender2021guide,pushkarna2022data}, there hasn't been a systematic analysis of the quality and informativeness of the model cards. In particular, we do not know what has and has not been documented. This is an important gap in our knowledge because adherence to community norms and documentation completeness directly impact the accessibility, transparency and user trust of AI models. As AI becomes increasingly ubiquitous in society, there is a need for regulation to keep up with the pace of technical progress~\cite{brundage2020toward}. Toward this end, industry-wide AI documentation standards can help to spur innovation and protect against harms~\cite{jobin2019global}, similar to how the electronics and automotive industries have converged toward universal technical specifications~\cite{mackiewicz2006overview}. Without comprehensive information on a model's functionality, limitations, and potential biases, users may not make informed decisions, leading to misuse or unintended consequences. Furthermore, without systematic analysis of current model cards, we risk propagating insufficient documentation practices, which can hinder efforts toward ensuring fairness, accountability, and equitable use of AI technologies.

To address this gap in knowledge, we present a comprehensive large-scale analysis on 32,111 AI model cards created from 6,392 distinct user accounts. These model cards are uploaded by developers to Hugging Face, which is one of the most popular model repositories for hosting cutting-edge AI models across a multitude of research domains and applications~\cite{alpaca-model,HuggingChat}. Distinct from GitHub, which hosts general software repositories, Hugging Face offers a user-friendly, AI-specific platform that provides an integrated machine learning pipeline. Importantly, it has established community guidelines for model cards, fostering a standardization that GitHub's user-dependent README documents lack, making it an ideal choice for our study. Through this analysis, we seek to characterize the extent to which the AI community has adopted and adapted model cards, the strengths and weaknesses of current documentation efforts, as well as to evaluate their impact on model development and usage. Our study reveals that, although the adoption of model cards has been largely successful, significant gaps persist in the completeness and quality of the provided documentation, with more than half of the models lacking a model card. The training section is most consistently completed, while sections addressing environmental impact, limitations, and evaluation exhibit the lowest completion rates, indicating a need for greater emphasis on these aspects of model documentation. 
Furthermore, our section content analysis of four key model card sections—Limitations, Uses, Evaluation, and Training—reveals an increased dialogue on data, often drawing comparable, if not surpassing, attention than the AI models themselves. 
To assess the impact of model cards on model utilization, we conducted an intervention study by adding detailed model cards to 42 popular models which have no or sparse model cards previously. This resulted in a moderate increase in weekly downloads, indicating the influence of well-documented model cards on model utilization. Our study provides a unique outlook on community norms and practices around model documentation through large-scale data science and linguistic analysis.

\section*{Results}

\subsection*{Data Overview of AI Models Hosted on Hugging Face}
Our analysis encompasses 74,970 AI model repositories on Hugging Face uploaded by 20,455 distinct user accounts as of October 1st, 2022. The number of models exhibits exponential growth, with a weekly growth rate of 3.16\% and a doubling time of 22 weeks (\textbf{Fig. \ref{fig:1}$a$}). As a sanity check, the number of model repositories reached 145,306 by March 4th, 2023, thereby confirming the exponential trend. The top 5 trending AI model\footnote{The trending models were defined as those receiving the most likes within a specific time period, following Hugging Face's official definition~\cite{huggingface-trending-models}.} repositories on Hugging Face over time reveal shifts in community attention (\textbf{Fig. \ref{fig:1}$b$}). Initially, GPT and BERT models dominated, reflecting a focus on language understanding and text generation. However, in Q3 2022, interest shifted toward image generative models, particularly Stable Diffusion and its variants, highlighting diverse interests beyond natural language processing.
Despite the widespread adoption of Hugging Face model cards, only 32,111 (44.2\%, contributed by 6,392 distinct user accounts) out of the 74,970 model repositories currently include model cards as Markdown README.md files within their model repo (\textbf{Fig. \ref{fig:1}$c$}). However, these models account for 90.5\% of total download traffic, highlighting the significance of model cards in fostering model adoption and usage. Here the download traffic is measured by the number of model repositories downloads. In light of these findings, our subsequent analyses will concentrate on the 32,111 models equipped with model cards.

\subsection*{Completeness of Model Cards Across Various Sections}

Model cards provides a standardized structure for conveying key information about AI models. Grounded in academic literature~\cite{modelcard} and official guidelines from Hugging Face~\cite{huggingface2021modelcard}, model cards conventionally comprise sections such as Training, Evaluation, Uses, Limitations, Environmental Impact, Citation, and How to Start. As illustrated in \textbf{Fig. \ref{fig:1}$d$}, these sections represent the essential constituents of a comprehensive model card. We parsed and evaluated the structure of model cards using a keyword-based detection pipeline for each section (e.g., detecting mentions of CO2 and its variants to identify the Environmental Impact section). 

Our evaluation indicates a significant disparity in community attention to different model card sections, a trend that seems to be expanding over time. Across all the model cards, Environmental Impact (2.0\%, or 639 out of 32,111 model cards), Citation (14.4\%), Evaluation (15.4\%), and Limitations (17.4\%) sections exhibit the lowest filled-out rates, while the Training section (74.3\%) is most frequently filled out (\textbf{Fig. \ref{fig:2}$a$}). Similar trends also hold for top 100 model cards and top 1000 model cards. Sections with lower fill-out rates also tends to have shorter average section word count among filled-out, indicating low community attention (\textbf{Fig. \ref{fig:2}$c$}). For example, the Environmental Impact section demonstrates both a low completion rate (3.7\%) among the top 1,000 model cards and a low average word count (68 words) among the filled-out ones. In contrast, the Training section exhibits the highest filled-out rate (71.0\%) and the second highest average word counts (168 words). Interestingly, despite its lower completion rate (22.1\%), the Limitations section tends to be the longest (151 words on average), hinting at the complex nature of discussing model limitations. Results are consistent across the top 100 model cards and the all the model cards (\textbf{Supp. Fig.~\ref{fig:S_word_count}}).

The disparities in community attention across different sections are progressively widening over time. A notable trend is the rapid increase in the filled-out rate for the Training section, which is increasing at 0.2\% per week (p = 3.31E-244), even after accounting for model categories such as tabular and natural language processing. The Environmental Impact section also exhibits an increase in its fill rate (p = 4.64E-24). An interesting discovery is that a large majority, about 84.8\% (542 out of 639), of the Environmental Impact sections appear to be automatically created by AI model-building tools~\cite{huggingface-co2emissions,huggingface-autotrain}, which not only make the AI model but also part of the model card. In particular, about 58.5\% (374 out of 639) of these sections mention "Model Trained Using AutoNLP," while around 26.3\% (168 out of 639) state "Model Trained Using AutoTrain." Moreover, their section text aligns perfectly with the template provided by AutoNLP/AutoTrain. The adoption of these automated tools to track CO2 emissions is a welcome change, as they increase awareness about the environmental impact of AI models. Through these tools, developers can better grasp the carbon footprint of their models, leading to more informed decisions during their creation and training process. Meanwhile, the filled-out rates for other sections are experiencing a decline (p < 0.001, \textbf{Supp. Table~\ref{tab:S-section-filled-out}}).

Top model cards associated with the most downloaded models are distinctive from an average model card in a number of ways. One distinctive feature is that they are considerably longer (\textbf{Fig. \ref{fig:2}$b$}, \textbf{Supp. Table~\ref{tab:S-word-count-top}}). The top 100 model cards are, on average, 1.35 times longer than the top 1,000 model cards (521 words vs. 384 words) and 2.73 times longer than an average model card at the population level (191 words). The section filled-out rate also differ significantly. Take top 100 model cards as an example. Although their Training section completion rate is similar to the population level (81\% vs. 74.3\%), they have significantly higher filled-out rates for Environmental Impact (9.0\% vs. 2.0\%), Limitations (39.0\% vs. 17.4\%), and Evaluation (47.0\% vs. 15.4\%) sections. The top 100 also have significantly higher filled-out rates in the Citation section (67.0\% vs. 14.4\%). These findings underscore the fact that top model cards are generally more detailed and structured, with a greater emphasis on sustainability and more thorough discussions of model performance and limitations. Furthermore, top models often have a strong connection to the academic research community, as shown by the high frequency of filled-out citation sections. Users may find these models more appealing because they come from scientific research, which is typically thoroughly checked for quality and accuracy by other experts in the same field.

\subsection*{A Deep Dive into Model Card Content}

To gain a comprehensive understanding of current practices and challenges in model documentation and identify areas for improvement, we conducted a content analysis of four critical model card sections on Hugging Face: Limitations, Uses, Evaluation, and Training. This analysis was motivated by Hugging Face's internal user study~\cite{huggingface-userstudies}, which identified the Limitations and Uses sections as the most challenging to write, and the fact that Evaluation and Training are two indispensable aspects of AI models. We employed a sentence-level topic modeling approach to accurately identify patterns and themes within the text, enabling a fine-grained analysis of model card sections compared to document-level topic modeling (see Methods). We quantified the prevalence of specific themes by calculating the frequency of sections containing sentences that mention those themes, both across all model cards and within the top 100 model cards.

\paragraph{Limitations Sections}
In the Limitations sections, our topic analysis uncovered a diverse array of subject matter, reflecting the myriad challenges and limitations faced by AI models. We identified three primary themes: disclaimers, data limitations, and model limitations. Disclaimers emerged in 11.6\% of the filled-out Limitations sections, often emphasizing the model is "not intended for production" or "should not be considered a clinical diagnostic tool," particularly for medical AI models. Explicit disclaimer sentences regarding third-party usage were also noted. In the top 100 model cards, a large majority, about 89.2\%, of the filled-out Limitations sections included such disclaimers. This stands in contrast to the typical practice in academic research papers, where disclaimers are less frequently seen~\cite{REAL-AI,limitations-are-not-properly-acknowledged-in-the-scientific-literature}. This distinction is probably a reflection of the low barrier of access and deployment of AI models on Hugging Face, thereby requiring explicit cautionary notes to mitigate risks tied to potential misuse of AI models. Data and model limitations received nearly equal attention, appearing in 30.1\% and 27.2\% of the filled-out Limitations sections, respectively. Their prevalence on the top 100 model cards are similar too. In data limitations, developers discussed the biases in training data, and the limited training data coverage. Discussions on model limitations revolved around both technical and societal aspects. From a technical standpoint, constraints, such as the maximum input length (e.g., 1024 tokens) for transformer-based models, were noted. On the societal front, concerns were raised about the biases in the AI model, as well as the potential risks that the AI model would inherit biases from its pre-trained backbone models (e.g., "since the model is further pretrained on the BERT model, it may have the same biases embedded within the original BERT model."). 

\paragraph{Uses Sections}
In the Uses sections, our topic analysis uncovered three primary themes: designated model functionality, operational guidelines, and misapplications. The most prominent theme in the Uses sections was the designated model functionality, featured in 58.2\% of the filled-out Uses sections. Here, developers clearly laid out the specific tasks the model was built for, such as "a fine-tuned model for generating tweets related to Indian politics." Closely linked was the theme of operational guidelines, covered in 25.2\% of the filled-out sections. This included practical information like installation steps, usage examples, checkpoint details, and fine-tuning instructions. Another important theme was misapplications, outlining the improper or out-of-scope use of the models. This theme, seen in 8.1\% of the filled-out sections, covered topics like malicious usage or the model's deployment in high-stakes situations. Although some overlap exists between the themes in the Limitations and Uses sections, each emphasizes distinct aspects. For example, broad statements on commercial usage might appear under a Disclaimer theme in Limitations, whereas specific warnings about misuse feature in the misapplication theme in Uses.

\paragraph{Evaluation Sections}
Our topic analysis of the Evaluation sections in model cards highlighted two key themes: evaluation data and evaluation results. Featured in 37.8\% of the filled-out sections, the first theme, evaluation data, described various datasets used for model testing. The second theme, evaluation results, appeared in 26.9\% of the filled-out sections, presenting performance metrics such as F1 scores and BLEU scores that depict a model's competence. Interestingly, these performance evaluations tend to present aggregate metrics across complete test datasets. Yet, such a broad approach can often obscure systematic errors within specific subgroups~\cite{daneshjou2022disparities}. For this reason, there's a growing call in the research community for more granular, subgroup-based evaluations that take into account factors such as age, gender, and location~\cite{Liang2022AdvancesCA,liangmetashift}. This fine-grained evaluation method aligns with practices in medicine. For instance, the U.S. Food and Drug Administration has mandated disaggregated clinical trial results~\cite{FDA1989}, requiring the analysis and reporting of outcomes by demographic subgroups, including sex, race, and age. This practice is designed to ensure the safety and efficacy of treatments across a diverse population and to uncover any disparities in response among different demographic groups. By adopting such an approach, emphasizing granular insights, the AI community can move towards a paradigm that promotes better understanding and communication of model strengths and limitations, driving towards a more inclusive and fair AI ecosystem, benefiting all users equally.

\paragraph{Training Sections}
In the Training sections, our analysis surfaced three main themes: hyperparameter configurations, training data, and training protocol. The theme of hyperparameter configurations, featured in 39.5\% of the completed sections, provides important information such as the number of epochs, batch size, and the selected optimizer. Equally important is the theme of training data, appearing in 32.7\% of the filled-out sections. This theme provides a comprehensive description not only of the volume and characteristics of the training dataset, but also of specific data pre-processing steps such as "convert all characters to lowercase." Recent studies have highlighted the time-intensive nature of data work in AI development~\cite{sambasivan2021everyone,anaconda2020}. The third theme, training protocol, appearing in 25.1\% of the filled-out sections, presents the technical steps required to replicate the training process. Taken together, these themes underscore a broader commitment within the AI community towards transparency and reproducibility by enabling researchers, practitioners, and industry stakeholders to efficiently build upon others' work~\cite{wolf2020transformers,mcmillan2021reusable}.

\subsection*{Model Card Intervention Study}
It is not well-understood how the quality of the model card affects model usage. To quantitatively study this relationship, we drew inspirations from well-established experimental design principles~\cite{campbell1963experimental} and designed a Model Card Intervention Study. 
We conducted an intervention study by adding detailed model cards to 42 popular models which has no or sparse model cards previously to evaluate the impact of model cards (\textbf{Fig. \ref{fig:4}$a$}). A Hugging Face employee with AI research background (one of the paper authors) wrote these model cards. The 42 models are sampled from model repositories that are (1) ranked top 5,000 based on cumulative downloads; (2) created in 2021 or 2022; and (3) had no model card or sparse model cards ($\le$ 100 words). It took on average 40 mins to create each detailed model card, which contains an average of 596.5 words. %
To ensure that the results are consistent across time, we upload the model card in two batches. The first batch ($N=26$) are uploaded within the week of 2022-11-07, and the second batch ($N=16$) are uploaded within the week of 2022-11-14. We then compared the average weekly downloads three weeks before and three weeks after the intervention. 

To control for shared temporal variation in downloads across models, we constructed a control group for both batches by randomly sampling $N=92$ models from the model repositories using the same filtering conditions (i.e. these control models also have no or similarly sparse model cards and have similar creation times and downloads as the models in the experimental group). To account for the unequal distribution of downloads across models, we normalize the weekly downloads for each model by dividing the corresponding average downloads before treatment. 
We then applied a difference-in-difference analysis~\cite{angrist2009mostly} to summarize the effect of the intervention: 
\begin{equation}
\label{eq:main-did}
    Y_{i,t} = \beta_0 + \beta_1 \cdot 1\{i=\text{Treatment}\} + \beta_2 \cdot 1\{t=\text{Post-intervention}\} + \beta_3 \cdot \{i=\text{Treatment}\}1\{t=\text{Post-intervention}\}.
\end{equation}
Here, $Y_{i,t}$ is the normalized downloads of a model in group $i$ (treatment or control) and time $t$ (pre- or post-intervention). In \eqref{eq:main-did}, $\beta_2$ and $\beta_2+\beta_3$ quantify the relative changes in downloads for the control and treatment groups, respectively. Therefore, $\beta_3$ (and the corresponding $p$-value for testing $H_0:\beta_3=0$) quantifies the impact of adding model cards on the weekly downloads.

\textbf{Fig. \ref{fig:4}$b$} shows the intervention results from the difference-in-difference analysis applied to Batches 1 and 2. For Batch 2, the model downloads in the treatment group significantly increased compared to the control group (p-value: 0.01). The average weekly downloads for the models in the treatment group increased by 29.0\% (95\% CI [10.6\%, 47.5\%]), while the control group did not see a significant change in downloads (p-value: 0.30). The results for Batch 1 were less clear: the average changes in downloads are 2.4\% (95\% CI [0.5\%, 4.4\%]) and 2.2\% (95\% CI [-1.5\%, 5.9\%]) for the control group and the treatment group, respectively. 
The observed effects might be limited due to external factors; for instance, the post-intervention period for Batch 1 fell with the 2022 Thanksgiving holiday season, which could have biased the estimated effect size towards the null, as users were less likely to download and use a model regardless of the model card quality. Additionally, the sample size used in the study is relatively small and could contribute to the differences in the results between the two batches. Further research with larger sample sizes and controlling for external factors may provide more conclusive insights into the relationship between model card quality and model usage.

\section*{Discussion}

In this paper, we present a comprehensive large-scale analysis on 32,111 AI model documentations on Hugging Face to understand the extent to which the AI community has adopted and adapted model cards, the strengths and weaknesses of current documentation efforts, as well as to evaluate their impact on model development and usage. Overall, our results indicate a widespread uptake of model cards within the AI community: 44.2\% of the models have a corresponding model card, and these models account for a significant 90.5\% of total download traffic. This substantial adoption underscores the community's recognition of the importance of model cards in facilitating model comprehension and application.

However, we discovered a significant dichotomy within the community's documentation practice. While model cards have been adopted on a broad scale, there's a striking disparity in the attention given to different sections of these cards. A majority of model cards incorporate a Training section, signifying a focus on model development. Conversely, sections detailing Environmental Impact and Evaluation are notably sparse, found in only 2.0\% and 15.4\% of model cards respectively. This gap extends to the Limitations section, which is only present in 17.4\% of the model cards. The growing divergence in the attention given to various sections, coupled with an increasing reluctance to address the limitations of models—both trends evidenced by our data— not only obstructs users from making informed decisions about model selection and usage but also undermines trust in these AI models. Our findings echo a recognized trend in the broader scientific community: the tendency to downplay study limitations, emphasizing successes while often overshadowing potential weaknesses and negative outcomes~\cite{REAL-AI, limitations-are-not-properly-acknowledged-in-the-scientific-literature}. Unlike traditional scientific papers which undergo peer review, model cards have no such mechanism for ensuring balanced and comprehensive documentation. This creates a unique challenge, indicating the need for novel solutions. Future work should focus on shaping strategies and standards to foster transparency and completeness in model card documentation. This is fundamental to building trust, advancing responsible AI use, and providing users with the critical information needed for informed model selection and application~\cite{Ethics-and-society-review-PNAS}.

Furthermore, our topic modeling analysis of four key model card sections—Limitations, Uses, Evaluation, and Training—reveals an emerging consensus regarding the vital role of data within the AI pipeline. In Limitation sections, data limitations and model limitations received nearly equal attention. In both Training and Evaluation sections, data emerges as a central theme. This emphasis on data in model documentation echoes existing literature that underscores the importance of data in AI model development. In practice, machine learning developers spend twice as much time on data as they do on the models themselves~\cite{sambasivan2021everyone,anaconda2020,kaggle2019}. Poor-quality data can significantly impact the performance, fairness, robustness, safety, and scalability of AI systems~\cite{halevy2009unreasonable,mehrabi2021survey}. Unfortunately, despite the criticality of data, it is widely acknowledged in the literature that data-related work is often undervalued in AI research. Data work is frequently perceived as “operational”~\cite{sculley2015hidden} compared to the more celebrated task of building novel models and algorithms. As a result, a data-centric focus is often lacking in current AI research. Recent calls have been made by researchers to prioritize data-centric AI research~\cite{Liang2022AdvancesCA}. Our results align with this trend and underscore the need for a more data-centric approach. By embracing and emphasizing the importance of data, the AI community can enhance the quality and reliability of models and foster advancements in responsible AI research.

In the model card intervention study, we experimentally examined the impact of detailed model cards on model usage. While the observed effect, a moderate increase in model downloads, suggests a positive influence of enhanced model cards, the findings should be interpreted with certain considerations. The timing of our post-intervention period for one batch coincided with the 2022 Thanksgiving holiday season, which might have led to an underestimation of the actual effect size due to generally reduced model activity. Furthermore, the sample size of 42 intervened models might have limited our statistical power to detect smaller effects. Despite these potential limitations, this study presents an important first step in quantitatively understanding the impact of model cards on model usage. In the broader context of open-source software documentation, previous studies have demonstrated that the organization of repository homepages and README files can significantly influence a project's perception, sustainability, and popularity, and even impact audience members' actions, such as joining the project~\cite{qiu2019signals,vasilescu2015gender,begel2013social,fan2021makes}. Given our preliminary results, we believe that it is worth pursuing more extensive, larger-scale randomized studies of model cards in the future. This can shed light on the nuanced ways in which model cards influence not just model downloads, but also broader aspects of model usage and their downstream impacts. The insights from such future work would be instrumental in guiding best practices for model card design and implementation, to fully realize their potential in promoting transparency, usability, and responsible AI practices.

Topics on how to create high-quality and responsible documentation should be incorporated into the AI curriculum~\cite{shen2021value,fiesler2020we,reich2020teaching,bates2020integrating,leidig2020acm}. Table~\ref{tab:1} presents a compilation of selected resources, including tools for generating model cards with graphical user interfaces, educational materials, and example model cards. These resources can serve as valuable teaching aids for students and practitioners alike. For example, green AI tools have been developed to automatically track model training CO2 emissions and document the environmental impact~\cite{huggingface-co2emissions,huggingface-autotrain}. Furthermore, the Hugging Face Model Cards Writing Tool simplifies the process of creating model cards by providing a user-friendly graphical interface, allowing individuals and teams with diverse skill sets and roles to collaborate effortlessly, even without coding or markdown knowledge. Documentation will be integral to the next stage of AI development, especially as we translate models from research sandbox to real-world deployment. By providing these resources, we aim to reduce barriers to entry and encourage the responsible deployment of AI models through transparent and accurate documentation.

\section*{Correspondence} 
Correspondence should be addressed to J.Z. (email: \href{mailto:jamesz@stanford.edu}{jamesz@stanford.edu}).

\section*{Competing interests} The authors declare no conflict of interest.

\section*{Acknowledgements} 
We thank D. McFarland and H. Fang for discussions. 
J.Z. is supported by the National Science Foundation (CCF 1763191 and CAREER 1942926), the US National Institutes of Health (P30AG059307 and U01MH098953) and grants from the Silicon Valley Foundation and the Chan-Zuckerberg Initiative.

\bibliography{main}

\appendix

\begin{figure}%
\centering
\begin{minipage}{1.00\textwidth}
    \centering
  \includegraphics[width=1.0\textwidth]{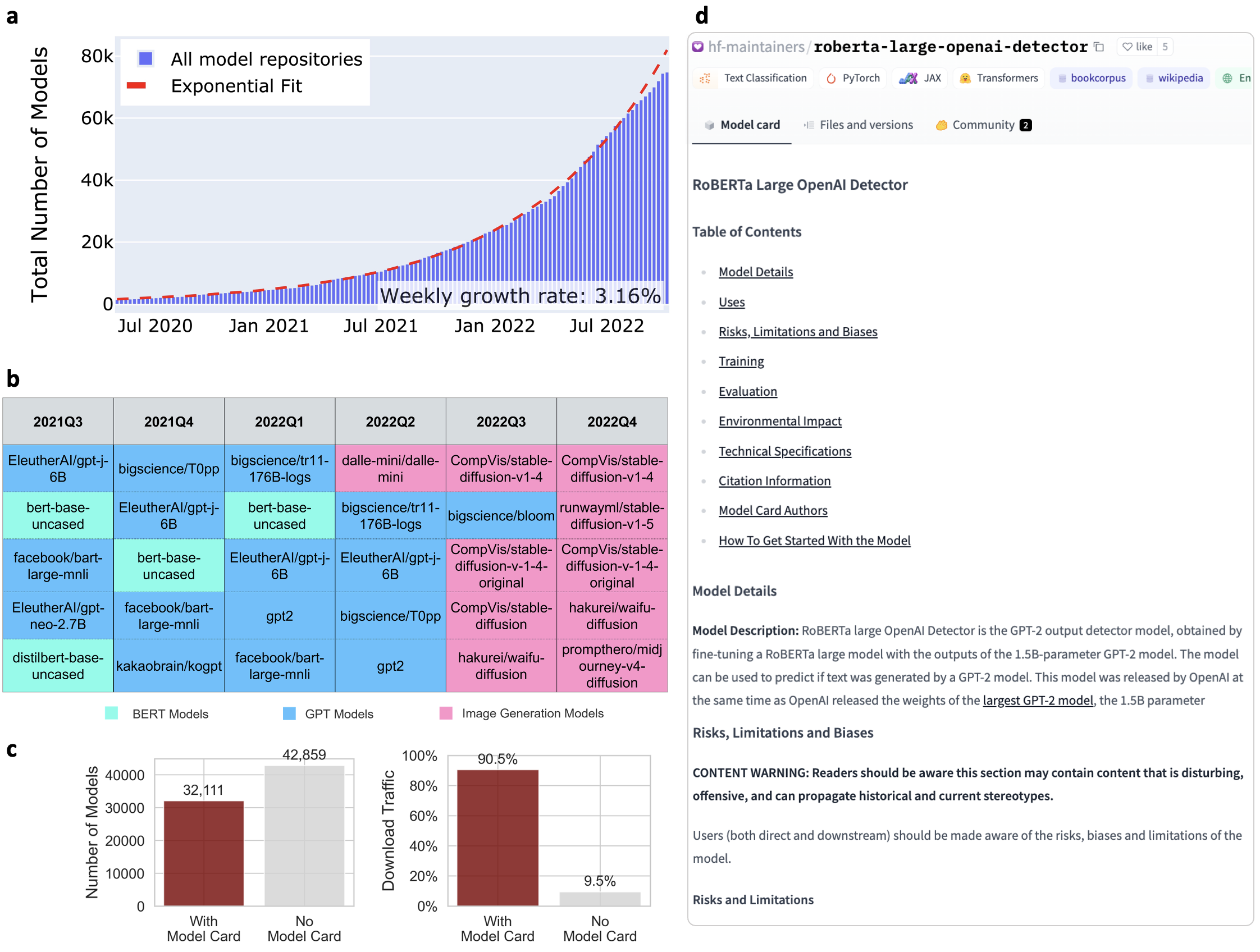}
\end{minipage}
\caption{
\textbf{Overview of 74,970 AI Models Hosted on Hugging Face.}
($\textbf{a}$) 
\textit{Exponential Growth of Hugging Face Model Repositories:} Our analysis indicates a substantial increase in the number of models hosted on Hugging Face, exhibiting a weekly growth rate of 3.16\% and a doubling time of 22 weeks. As of October 1st, 2022, there are 74,970 AI model repositories available on the platform.
($\textbf{b}$)
\textit{Temporal Trends in Model Popularity:} Examination of the top-5 trending AI model repositories on Hugging Face reveals dynamic shifts in community focus towards various model types. Language understanding and text generation models initially dominate, while image generative models gain prominence in later stages.
($\textbf{c}$)
\textit{Model Card Adoption and Download Traffic:} Despite only 42.8\% of model repositories (32,111 out of 74,970) featuring model cards, these models account for an overwhelming 90.5\% of total download traffic on the platform.
Our analysis focuses on these 32,111 models for further examination.
($\textbf{d}$) 
\textit{Model Card Illustration:} An exemplary model card is displayed, showcasing the essential components and information typically provided by a model card. 
}
\label{fig:1}
\end{figure}

\begin{figure}%
\centering
\begin{minipage}{1.00\textwidth}
    \centering
  \includegraphics[width=0.95\textwidth]{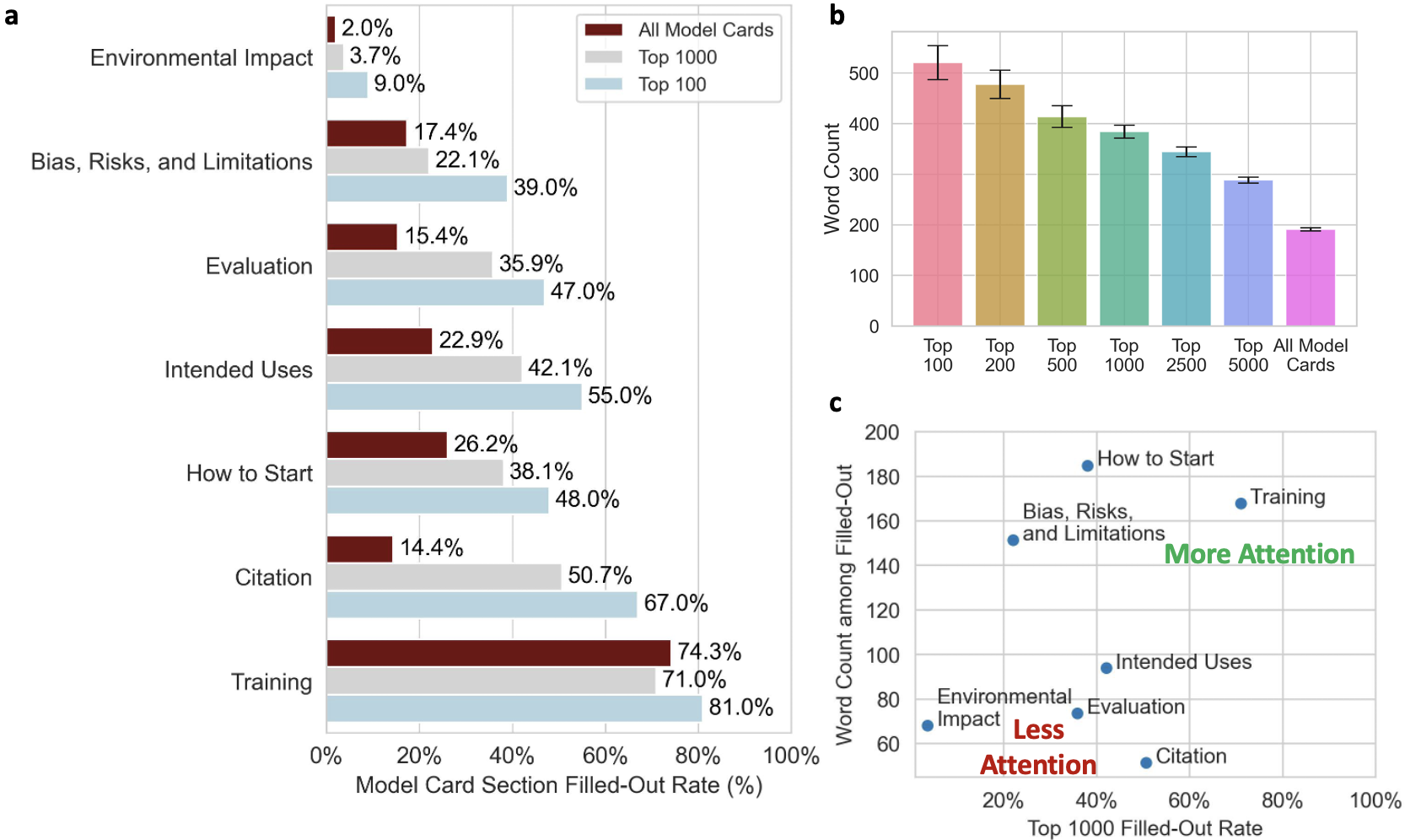}
\end{minipage}
\caption{
\textbf{Section Analysis of Hugging Face Model Cards.}
($\textbf{a}$) 
\textit{Low Filled-out Rates for Model Card Sections:} This panel presents the filled-out percentages for each section in 32,111 model cards, as well as the top 100 and top 1,000 model cards ranked by downloads. Environmental Impact, Limitations, and Evaluation sections exhibit the lowest filled-out rates, while the training section is most frequently filled out.
($\textbf{b}$) 
\textit{Highly Downloaded Models Have Longer Model Cards.} 
Model cards in the top tier, ranked by downloads, are notably longer, suggesting a positive correlation between model card length and their usage. The total word count across all the sections is shown on the y-axis. 
($\textbf{c}$) 
\textit{Disparate Community Attention Patterns Across Model Card Sections:} The Environmental Impact section demonstrates both a low filled-out rate among the top 1,000 model cards and a low average word count, indicating low community attention. Conversely, the Training section exhibits high filled-out rates and average word counts, signifying more community attention.
}
\label{fig:2}
\end{figure}

\begin{figure}%
\centering
\begin{minipage}{1.00\textwidth}
    \centering
  \includegraphics[width=\textwidth]{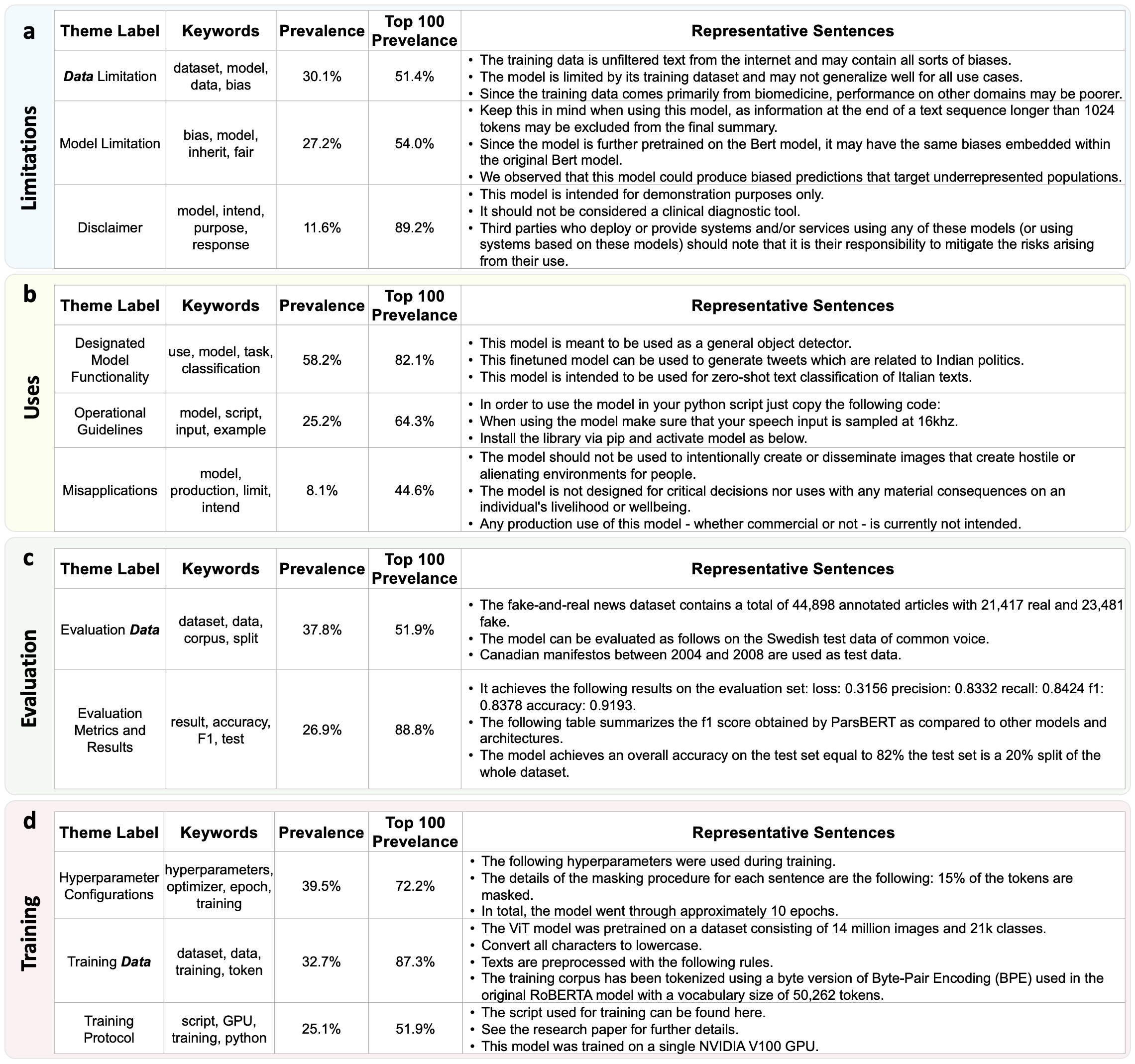}
\end{minipage}
\caption{
\textbf{Uncovering Key Themes in Model Card Sections through Topic Modeling Analysis.}
This figure displays the outcomes of our topic modeling assessment on the contents of the ($\textbf{a}$) Limitations Section, ($\textbf{b}$) Uses Section, ($\textbf{c}$) Evaluation Section, and ($\textbf{d}$) Training Section of model cards. Each panel illustrates the human-assigned theme label, prevalence, associated keywords, and representative sentences for each section. 
Prevalence reflects the frequency of sections containing sentences mentioning a specific theme, and top 100 prevalence is the prevalance measured on only the top 100 model cards.  
The keywords and representative sentences were algorithmically determined. 
Theme labels were manually assigned based on the grouping of topic modeling sentence clusters. Our section content analysis identifies practitioners' priorities within each section, highlighting the community's focus on discussing data, sometimes with equal or even greater emphasis than the model itself. 
}
\label{fig:3}
\end{figure}

\begin{figure}%
\centering
\begin{minipage}{1.00\textwidth}
    \centering
  \includegraphics[width=0.5\textwidth]{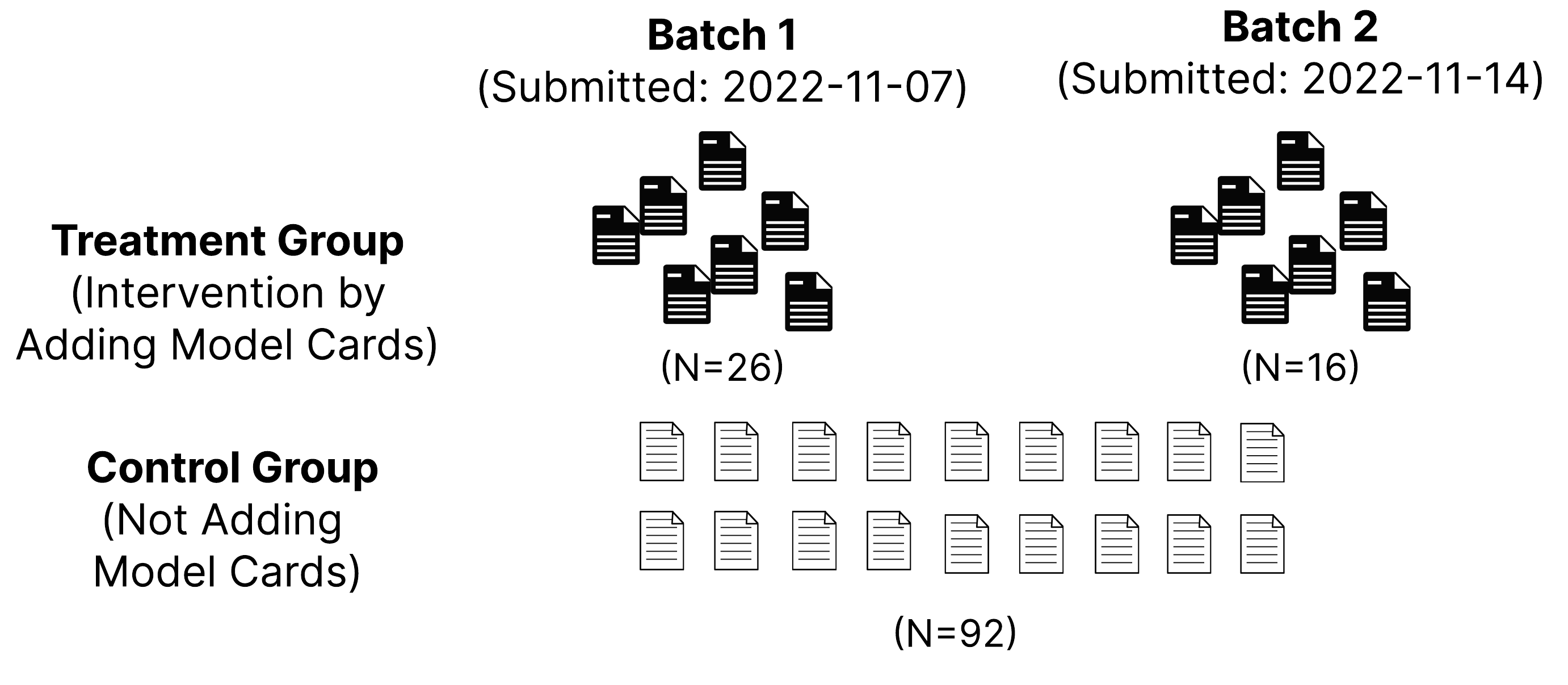}
  \includegraphics[width=0.35\textwidth]{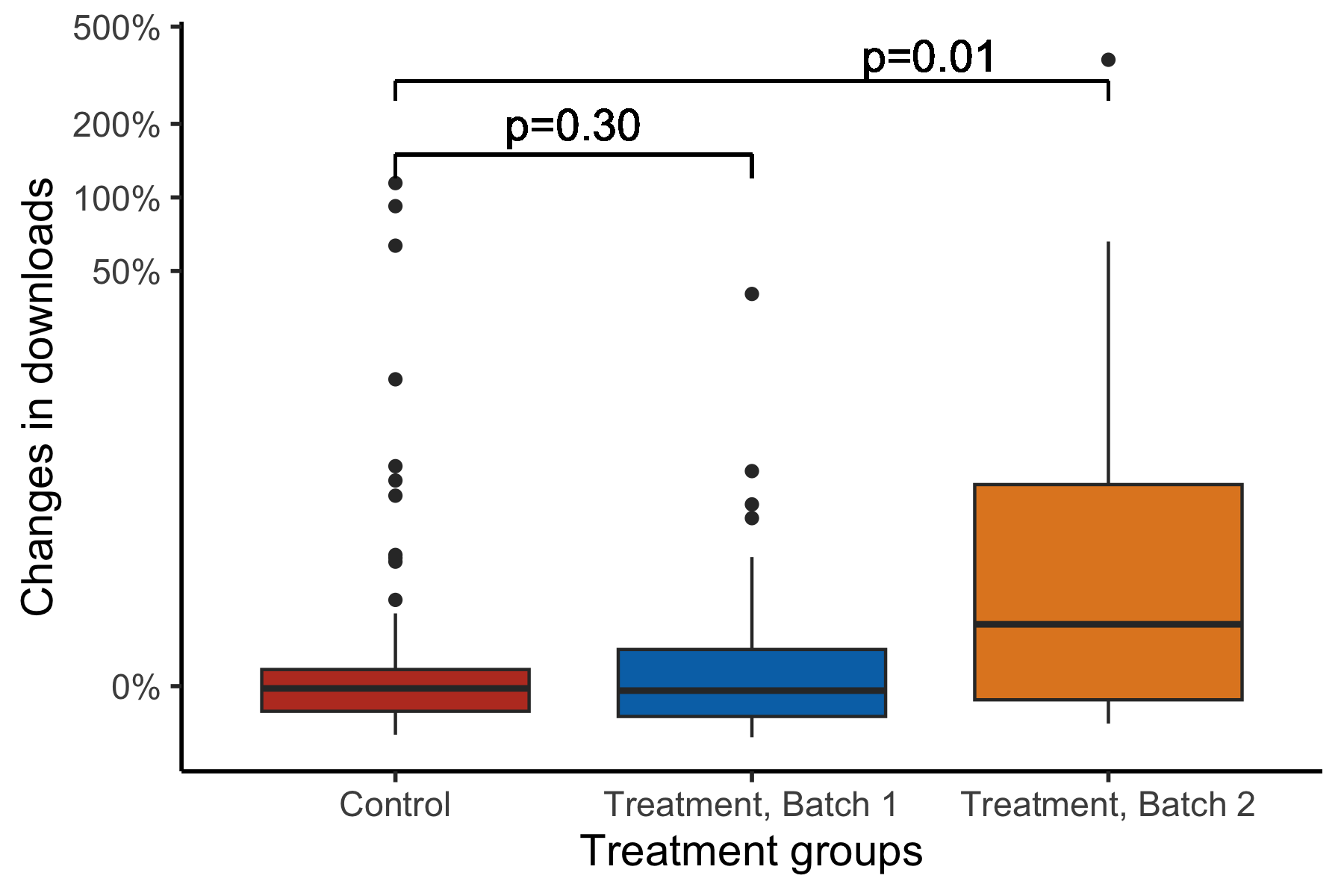}
\end{minipage}
\begin{centering}
\hspace{35mm} (a) \hspace{75mm} (b) \\
\end{centering}
\caption{
\textbf{Model Card Intervention Study.}
($\textbf{a}$) 
Experimental design: A schematic representation of the Model Card Intervention Study, delineating the selection of models, division into treatment (two batches) and control groups, and model card intervention process.
($\textbf{b}$) 
Outcome: Box plots displaying the percentage change in average weekly downloads for the treatment and control groups in Batches 1 and 2. For each color-filled box, three horizontal lines correspond to the 25th, 50th, and 75th percentiles; the upper (lower) whiskers extend from the 75th (25th) percentiles to the largest (smallest) value no further than 1.5 $\times$ interquartile range. Statistical significance ($p$-values) is indicated for each batch.
Overall, our analysis revealed a moderate effect of model cards on model downloads. 
}
\label{fig:4}
\end{figure}

\clearpage
\newpage
\begin{table}%
\centering
\includegraphics[width=\textwidth]{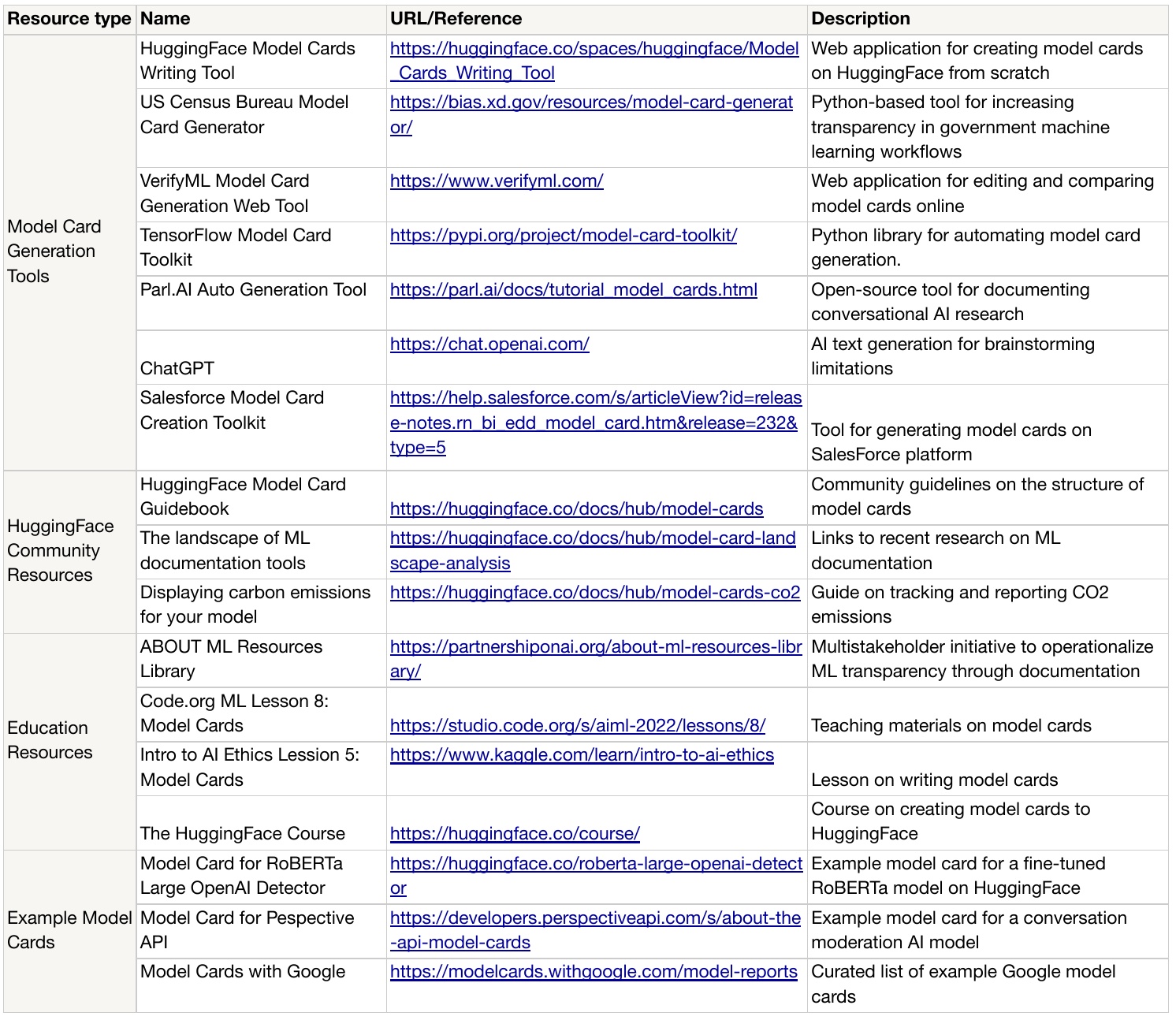} %
\caption{Selected resources for promoting responsible and informative model card creation.}
\label{tab:1}
\end{table}

\newcounter{suppfigure}
\newenvironment{suppfigure}[1][]{%
  \addtocounter{suppfigure}{1}%
  \renewcommand{\thefigure}{S\arabic{suppfigure}}%
  \begin{figure}[#1]%
}{%
  \end{figure}%
}
\newcounter{supptable}
\newenvironment{supptable}[1][]{%
  \addtocounter{supptable}{1}%
  \renewcommand{\thetable}{S\arabic{supptable}}%
  \begin{table}[#1]%
}{%
  \end{table}%
}

\clearpage
\newpage

\section*{Materials and Methods}
\subsection*{Model Cards as AI Documentation Framework}

A plethora of documentation tools have been proposed to address different aspects of AI systems, with early efforts focused on data and later expanding to capture the AI development lifecycle more comprehensively. Data-focused documentation tools~\cite{bender2018data,bender2021guide,pushkarna2022data}, including datasheets for datasets~\cite{datasheets} and data nutrition labels~\cite{holland2018dataset,chmielinski2022dataset}, serve as a vehicle for communication between dataset creators and consumers, addressing the lack of industry standards for documenting AI datasets. These tools document the contexts and contents of datasets, including their motivation, composition, collection process, and recommended uses, facilitating greater transparency, accountability, and reproducibility of AI results while mitigating unwanted biases in AI models. They also enable dataset creators to be intentional throughout the dataset creation process. Datasheets and other forms of data documentation are increasingly released along with datasets, helping researchers and practitioners to choose the right dataset for their needs. 
These tools emphasize the data lifecycle~\cite{hutchinson2021towards}, encompassing aspects like assembly, collection, and annotation. Later, efforts expanded to capture the AI development lifecycle more comprehensively, emphasizing the importance of an iterative design process to ensure accessibility to users with diverse backgrounds and goals when interacting with model cards – including developers, students, policymakers, ethicists, those impacted by AI models, and other stakeholders. Model cards~\cite{modelcard} and FactSheets~\cite{FactSheets} serve as reporting tools for trained AI models, documenting evaluation, usage, and other relevant aspects. Model cards foster transparency, accountability, comparability, and user confidence, which are essential for responsible development and deployment of AI models~\cite{transparency1,raji2019ml,Accountability1, Accountability2,zou2018design,kulesza2015principles}. Furthermore, model cards also have the potential to assist in regulatory compliance by offering a structured framework for documenting and communicating key information about a model's performance, training, and evaluation process~\cite{FactSheets,seifert2019towards,regulation2016regulation,goodman2017european}. Value cards~\cite{shen2021value} educate students and practitioners about values related to AI models, while consumer labels for AI models~\cite{seifert2019towards} help non-experts understand key concerns. Researcher are also exploring the design of system cards for complex systems such as AI-based content ranking which often involves a pipeline of many AI models, and discussing complicated issues including use policies, access controls, and monitoring for abuse. 
In the broader context of open-source software documentation, previous studies have demonstrated that the organization of repository homepages and README files can significantly influence a project's perception and even impact audience members' actions, such as joining the project~\cite{qiu2019signals,vasilescu2015gender}. Well-structured READMEs have been linked to enhanced sustainability and popularity of projects~\cite{begel2013social,fan2021makes}. These insights underscore the necessity for comprehensive and user-friendly documentation to promote user trust, engagement, and adoption of AI models and tools.

\subsection*{Accessing and Parsing Model Cards}

We employed the Hugging Face's library, $huggingface\_hub$, for accessing the model cards on the Hugging Face Hub. 
The $huggingface\_hub.list\_models()$ API grants access to all models hosted on the Hub, which numbered 74,970 as of October 1st, 2022. Additionally, the library's $ModelCard$ module allows for the loading of existing cards from the hub. By examining the $CardData$ metadata of models, we can verify the presence of a model card, and employ $huggingface\_hub.ModelCards.load()$ function to obtain the card associated with a given model.

Model cards are written in Markdown format and serve as the README file for the model repository. To extract plain descriptive text for further analysis, we utilized the Python package mistune (\url{https://mistune.readthedocs.io/en/latest/}) to parse the README file and remove features such as tables, codes, links, and images. The structure of model cards plays a crucial role in conveying key information. Drawing from academic literature~\cite{modelcard} and Hugging Face's official guidelines, model cards typically include sections such as Training, Evaluation, Uses, Limitations, Environmental Impact, Citation, and How to Start. An example model card, shown in \textbf{Fig. \ref{fig:1}$d$}, demonstrates the essential components and information provided by model cards.

However, due to the subjective and dynamic nature of section naming, sections that discuss the same topic may not have the same title. For example, both the "Training Procedure" and "Training Results" sections may address aspects of the training. To categorize sections with the same topic, we use a keyword-based detection method to identify similar topics. We define categories based on Hugging Face Model Card template.
We then applied keyword detection on both the section titles and contents to categorize the sections. It is important to note that section detection could not be solely based on the section titles. While some section titles, such as "Training Data," may contain keywords that allow direct categorization, other information may be hidden within the section contents. For instance, in the "Model Trained Using AutoTrain" section, the topic being discussed cannot be inferred from the section title alone, but the section contains information about the amount of CO2 generated by AutoTrain, which can be categorized as Environmental Impact. By utilizing keyword detection, we achieved a coverage of categorized sections of 89.51\%, which validates the reliability of our categorizing method.

\subsection*{Topic Modeling}

We leverage a sentence-level topic modeling approach to identify the patterns and themes within the section. For each section, we first employ NLP methods to split it into sentences and label the sentences with their original section, which constructs a sentence-section map. Then we deploy BERTopic\cite{grootendorst2022bertopic}(\url{https://maartengr.github.io/BERTopic/}) approach to automatically cluster the themes. BERTopic is a topic modeling technique that utilizes BERT\cite{devlin2018bert} embeddings and class-based TF-IDF (c-TF-IDF) to develop compact and coherent clusters, thereby enabling the generation of easily understandable topics while preserving salient terms within the topic descriptions. It starts by converting the documents to numerical representations. The sentences will be embedded using the default embedding model Sentence Transformer\cite{reimers-2020-multilingual-sentence-bert} all-MiniLM-L6-v2. Then Uniform Manifold Approximation and Projection\cite{2018arXivUMAP} (UMAP) is used to reduce the dimensionality of document embedding into something easier for clustering. After reducing the embeddings, it starts clustering the data. For that, a density-based clustering technique, Hierarchical Density-Based Spatial Clustering of Applications with Noise\cite{mcinnes2017accelerated} (HDBSCAN) is used to cluster the data. It can find clusters of different shapes and has the nice feature of identifying outliers where possible. When using HDBSCAN to do clustering, the bag-of-words representation is used on the cluster level for the purpose of topic representation, in which the frequency of each cluster can be found. That is, all documents in a cluster can be combined into a single document, and we can count how often each word appears in each cluster. From the generated bag-of-words representations, c-TF-IDF is used to extract the most important words per cluster. 
$$\mathbf{W}_{x, c}=\left\|\mathbf{tf}_{x, c}\right\| \times \log \left(\mathbf{1}+\frac{\mathbf{A}}{\mathbf{f}_{x}}\right),$$

where $\mathbf{W}_{x, c}$ is the c-TF-IDF score of word $x$ in class $c$, $\mathbf{tf}_{x, c}$ is the frequency of word $x$ in class $c$, $\mathbf{f}_{x}$ is the frequency of word $x$ accross all classes, and $\mathbf{A}$ is the average number of words per class. Then the words are sorted based on their c-TF-IDF score, and the keywords are the top words. The representative sentences are selected randomly in each cluster.

By employing this, we have got the automatically clustered topics generated from BERTopic with their keywords and representative sentences. However, the clustered topics are too fine-grid, and we identify some topics that are acceptable to be merged. For example, in the training section, topic $2\_used\_dataset\_containing\_commoncrawl$ and topic $4\_bookcorpus\_books\_11038\_tables$ are both about the training data. Although we can fine-tune the parameters of BERTopic to get better results, the fine-tuning process could be time-consuming and might miss some important patterns. Therefore, we manually identify similar clustered topics based on the keywords and representative sentences, and use BERTopic to merge them and get the new keywords and representative sentences.

Finally, we use the sentence-section map constructed previously and map the sentences to sections, and calculate the size of sections within each topic. This allows us to identify the most prominent themes in each section and the size of each theme relative to the section.

\subsection*{Code and Data Availability}
The analysis code and the set of 42 model cards created in this study are publicly available on GitHub \url{https://github.com/Weixin-Liang/AI-model-card-analysis-HuggingFace}. The Hugging Face model cards data can be accessed through the Hugging Face Hub API at \url{https://huggingface.co/docs/huggingface_hub/package_reference/hf_api}.

\begin{suppfigure}[tb]%
\centering
\begin{minipage}{0.95\textwidth}
    \centering
  \includegraphics[width=0.95\textwidth]{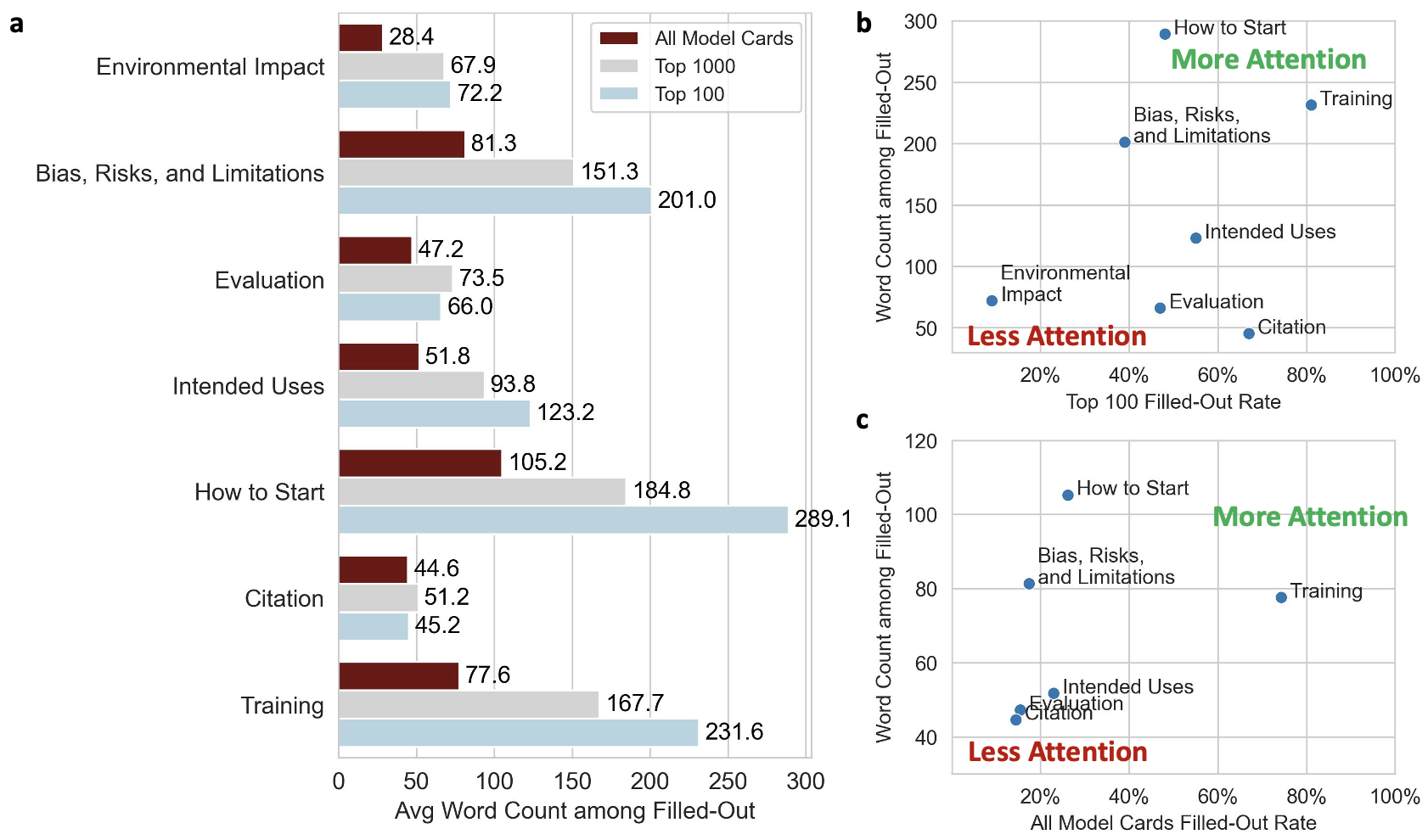}
\end{minipage}
\caption{
\textbf{In-depth Analysis of Section Word Counts in Model Cards.}
($\textbf{a}$) 
\textit{Comparative Assessment of Average Section Lengths in Model Cards Based on Word Count.}
This figure displays the average section length, measured in word count, among completed sections for all model cards, the top 1000 model cards, and the top 100 model cards. Sections such as How to Start, Training, and Limitations are substantially longer, while Citation, Evaluation, Environmental Impact, and Intended Uses are relatively shorter. Interestingly, despite its lower completion rate, the Limitations section exhibits one of the highest average word counts (161 words in the top 1000 model cards).
($\textbf{b-c}$) 
\textit{Disparate Community Attention Patterns Across Model Card Sections, Analyzed for both the top 100 model cards ($\textbf{b}$) and all model cards ($\textbf{c}$).} The Environmental Impact section demonstrates both a low completion rate and a low average word count, indicating limited community attention. In contrast, the Training section displays high completion rates and average word counts, signifying greater community engagement.
}
\label{fig:S_word_count}
\end{suppfigure}

\begin{supptable}[ht]
  \centering
  \includegraphics[width=\linewidth]{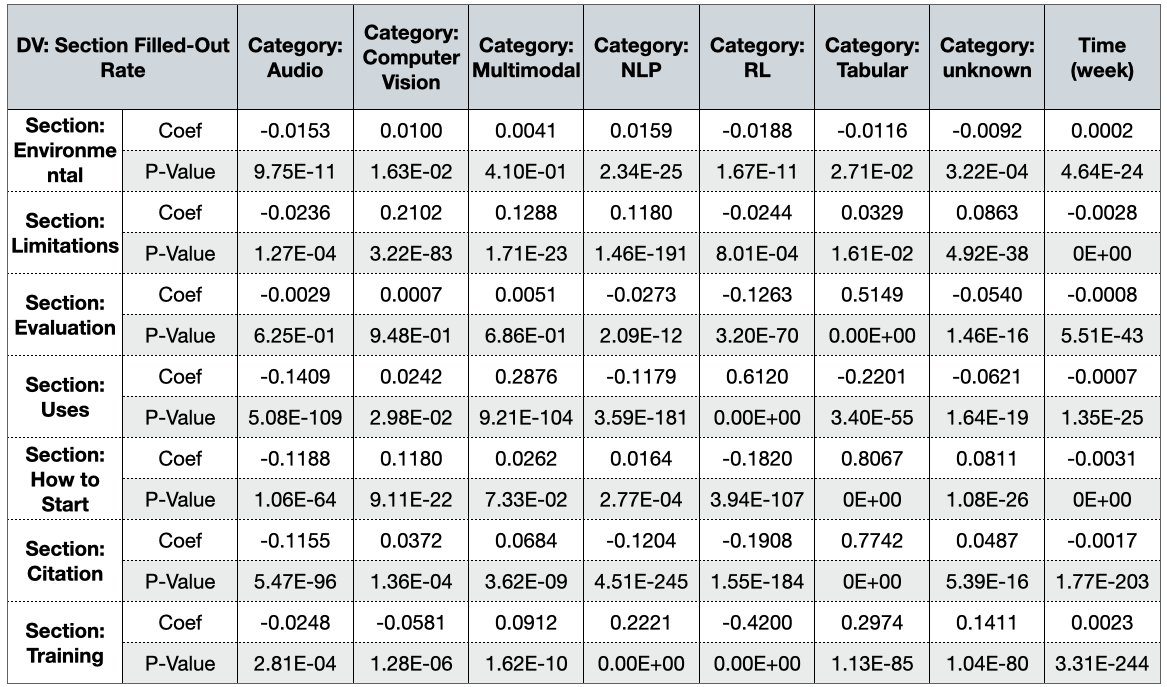}
  \caption{
  \textbf{Temporal trends in the completion rates of different model card sections.} 
  The table displays the results of a linear regression analysis conducted to assess the changes in the fill rates of different sections of model cards over time, while controlling for model categories (e.g., tabular and natural language processing). A significant trend is the swift increase in the completion rate for the Training section, rising at 0.2\% per week (p = 3.31E-244). In contrast, most other sections are experiencing a decline in their completion rates (p < 0.001). An interesting exception to this pattern is the Environmental Impact section, which demonstrates a rise in its completion rate (p = 4.64E-24). 
  }
  \label{tab:S-section-filled-out}
\end{supptable}

\begin{supptable}[ht]
  \centering
  \includegraphics[width=\linewidth]{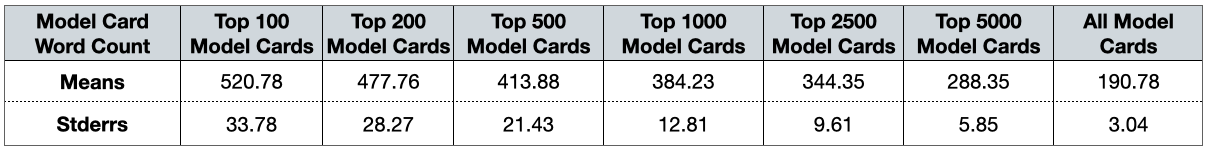}
\caption{
\textbf{Highly Downloaded Models Have Longer Model Cards.} 
Model cards in the top tier, ranked by downloads, are notably longer, suggesting a positive correlation between model card length and their usage. The total word count across all the sections is shown on the y-axis. }
  \label{tab:S-word-count-top}
\end{supptable}

\end{document}